\documentclass[iop,superscriptaddress]{emulateapj}

\usepackage{graphicx}
\usepackage{dcolumn}
\usepackage{amssymb} 
\usepackage{ctable}
\usepackage{bm}
\usepackage{epsfig}
\usepackage{epstopdf}
\usepackage{epsf,color}
\usepackage{natbib}

\newcommand{\cmnt}[1]{}

\def\vecx{\mathbf{x}}

\def\VEV#1{\left\langle #1 \right\rangle}

\def\veca{\mathbf{a}}
\def\vecg{\mathbf{g}}
\def\vecA{\mathbf{A}}

\begin{document}

\title{Nonlinear evolution of dark matter subhalos and applications to warm dark matter}

\author{Anthony R. Pullen}
\affiliation{Jet Propulsion Laboratory, California Institute of Technology, 4800 Oak Grove Drive, MS 169-237, Pasadena, CA, 91109, U.S.A.}
\author{Andrew J. Benson}
\affiliation{Carnegie Observatories, 813 Santa Barbara Street, Pasadena, CA 91101, U.S.A}
\author{Leonidas A. Moustakas}
\affiliation{Jet Propulsion Laboratory, California Institute of Technology, 4800 Oak Grove Drive, MS 169-506, Pasadena, CA, 91109, U.S.A.}

\email{anthony.r.pullen@jpl.nasa.gov}

\begin{abstract}
We describe the methodology to include nonlinear evolution, including tidal effects, in the computation of subhalo distribution properties  in both cold (CDM) and warm (WDM) dark matter universes.  Using semi-analytic modeling, we include effects from dynamical friction, tidal stripping, and tidal heating, allowing us to dynamically evolve the subhalo distribution.  We calibrate our nonlinear evolution scheme to the CDM subhalo mass function in the Aquarius N-body simulation, producing a subhalo mass function within the range of  simulations.  We find tidal effects to be the dominant mechanism of nonlinear evolution in the subhalo population.  Finally, we compute the subhalo mass function for $m_\chi=1.5$ keV WDM including the effects of nonlinear evolution, and compare radial number densities and mass density profiles of subhalos in CDM and WDM models.  We show that all three signatures differ between the two dark matter models, suggesting that probes of substructure may be able to differentiate between them.
\end{abstract}
\keywords{cosmology: theory; dark matter; galaxies: formation; galaxies: halos}
\maketitle

\section{Introduction}
The behavior of dark matter on the largest scales is described precisely by the cold dark matter (CDM) paradigm \citep{1982ApJ...263L...1P}, as evidenced in cosmic microwave background (CMB) and large-scale structure (LSS) observations \citep{2011A&A...536A...1P,2013arXiv1303.5062P,2012MNRAS.427.3435A}. However, on collapsed dark-matter-halo scales, departures from the CDM paradigm are possible, and even desirable.  Several alternative dark matter (DM) models exist in the literature \citep{2004ASSL..301..263M,2005MNRAS.363.1092A,2005A&A...438..419B,2007MNRAS.382.1225M,2008ApJ...679.1173R,2009JCAP...05..012B,2012MNRAS.420.2318L}, as solutions to the core-vs.-cusp problem \citep{2001MNRAS.320L...1S,2004MNRAS.353L..17D,2009MNRAS.397.1169D,2009ApJ...706.1078N,2011ApJ...728L..39N,2010AdAst2010E...5D,2011ApJ...741L..29K,2011MNRAS.414.3617K,2012MNRAS.420.2034S,2012arXiv1203.4240W} and the missing satellite problem \citep{2011MNRAS.415L..40B,2012MNRAS.422.1203B,2012MNRAS.424.2715W}, but these observations may be the result of galaxy formation physics that is not yet well understood \citep{2002MNRAS.333..177B,2007MNRAS.374...16L,2009MNRAS.397L..87L,2010MNRAS.404.1129S,2011MNRAS.417.1260F,2011AJ....142...24O,2012MNRAS.422.1231G,2012ApJ...749...36K,2012ApJ...752...24P,2012MNRAS.421.3464P,2013MNRAS.429..725S,2013MNRAS.428.1696V}.  Various experimental measurements should be sensitive to dark matter phenomenology on small scales \citep{2005ApJ...621..757S,2008PhRvL.100d1304V,2010MNRAS.407..225V}, in particular future lensing experiments \citep{2009ApJ...699.1720K,2009MNRAS.392..945V,2009MNRAS.400.1583V,2010MNRAS.408.1969V,2012Natur.481..341V}.

The subhalo mass function has arisen as a potential tracer of dark matter phenomenology and clustering properties \citep{2010PhRvD..82l3521P,2012PhRvD..85d3514W,2013PhRvD..88l3515W,2014MNRAS.442.2487K,2014PASA...31....6M,2013arXiv1312.3325C,2014PhRvL.112p1303A}.  In order to use subhalo measurements to constrain dark matter properties, halo population evolution must be rapidly and accurately predicted for theoretical DM models.  \citet{2013MNRAS.428.1774B} (hereafter Paper I) recently used a semi-analytic galaxy-formation code to construct halo mass functions for warm dark matter (WDM) by implementing a generalized extended Press-Schecter formalism, with an mass-scale dependent excursion set barrier for collapse.  This code, \textsc{Galacticus}, is highly modularized, and available publically \citep{2012NewA...17..175B}.\footnote{https://sites.google.com/site/galacticusmodel/home}  In Paper I, merging times for subhalos were not considered since only halo mass functions were computed, and tidal effects were neglected. Tidal stripping suppresses halo abundances, particularly at small mass scales, where WDM halos are also suppressed.  To make predictions for different DM models, nonlinear evolution must be properly taken into account. These processes are typically included in semi-analytic models via the use of fitting formulae. For example, \citet{2008ApJ...675.1095J} provides a fitting formula calibrated to N-body simulations for the merging times of subhalos based on orbital parameters. Similarly, \citet{2005MNRAS.359.1029V} provides a fitting function for mass loss rates from subhalos, but it too is calibrated only for CDM simulations. Since we wish to explore the effects of these processes in WDM universes, a fitting formula approach cannot be employed. We will compare some of our CDM results to those obtained using a ``simple'' model in which we instead employ the fitting formula of \citet{2008ApJ...675.1095J}.

In this paper, we include the effects of dynamical friction, tidal stripping, and tidal heating in semi-analytic galaxy formation calculations to track the evolution of the subhalo populations of CDM and WDM halos.  In our code we dynamically evolve the orbits of the subhalos, calculating at each timestep the position, velocity, and mass of each subhalo.  We perform this analysis for Milky-Way-sized halos, though this is extendable to other scales.  We also determine the expansion of subhalos due to tidal heating, which accelerates the mass loss.  The parameters describing the tidal effects are set by calibrating the subhalo mass function for CDM to the Aquarius simulation \citep{2008MNRAS.391.1685S}.  We also separate the effects of the different nonlinear evolution mechanisms on the subhalo mass function, showing that tidal effects have the dominant influence on the mass function.  We then compute the WDM mass function for a test case using the Paper I WDM formalism but including nonlinear evolution.  Finally, we compare the CDM and WDM subhalo mass functions, radial subhalo distributions, and subhalo density profiles.  

The plan of our paper is as follows:  we describe the implementation of nonlinear evolution in our semi-analytic modeling, the calibration of the CDM model to simulations, and the implemented changes for the WDM model, in Sec.~\ref{S:nonlinevo}.  In Sec.~\ref{S:cdmwdm} compare the CDM results with those from WDM.  We conclude in Sec.~\ref{S:conc}.  Wherever not explicitly mentioned, we assume a flat $\Lambda$CDM cosmology with parameters compatible with WMAP7 \citep{2011ApJS..192...18K}, namely $\Omega_m=0.2725$, $\Omega_\Lambda=0.7275$, $\sigma_8=0.807$, $n_s=0.961$, and $h=0.702$.

The methods described in Sec.~\ref{S:nonlinevo} have been implemented within the open source semi-analytic galaxy formation code, \textsc{galacticus} \citep{2012NewA...17..175B}.  All results presented in this section were generated using \textsc{galacticus} v0.9.3.  Control files and scripts to generate all results presented in this paper using  \textsc{galacticus} can be found at https://sites.google.com/site/galacticusmodel/home.

\section{Nonlinear Evolution Theory} \label{S:nonlinevo}
In this section we determine the various nonlinear evolution effects we include in our semi-analytic calculation.  Much of this work is based on previous work in the literature, in particular \citet{2001ApJ...559..716T}, \citet{2002MNRAS.333..156B}, and \citet{2005ApJ...624..505Z}.  To implement these effects properly, we evolve the position and velocity vectors of each subhalo according to interactions with the parent halo, neglecting interactions between subhalos.\footnote{While we do not include subhalo-subhalo interactions explicity, we do account for the mass in subhalos by including subhalo masses in the mass of the host halo.  Specificaly, we make the approximation that we can compute the evolution of one subhalo by smoothing the mass of all other subhalos over the host halo (such that the mass in subhalos is distributed similarly to the mass in the host halo).  \cite{2005MNRAS.364..977P} explores how this picture changes if subhalo-subhalo interactions are explicitly taken into account, showing that this type of interaction is negligible.}  The acceleration $\veca_{\rm sat}$ of a subhalo (or satellite) can be written as
\begin{eqnarray}
\veca_{\rm sat} = -\vecg + \veca_{\rm df}\, ,
\end{eqnarray}
where $\vecg$ is the host halo potential gradient, and $\veca_{\rm df}$ is the deceleration due to dynamical friction, which causes subhalos to fall into and merge with the host halo.  Note that we assume the host halo is static.\footnote{By a \emph{static} halo, we mean one which does not respond to the presence of a subhalo (we add in that effect via the Chandrasekhar dynamical friction formula).  The halo is not static as a function of time -- it will grow at a cosmological rate as determined from our merger trees.} We evolve the subhalo orbit until it merges when passing too close to the parent halo or loses too much mass due to tidal stripping such that the subhalo is no longer distinct.  Mass loss can also affect the dynamics in that deceleration due to dynamical friction being dependent on the subhalo mass.  Thus, we present the formalism for these effects.

\subsection{Dynamical Friction}
Dynamical friction decelerates the motion of satellites relative to the host halo through gravitational interactions.  To implement dynamical friction in our calculation, we use the Chandrasekhar formula \citep{1943ApJ....97..255C}
\begin{eqnarray} \label{E:dynfric}
\veca_{\rm df} &=& -4\pi G^2\ln\Lambda M_{\rm sat}\rho_{\rm host}(r_{\rm sat})\frac{\mathbf{V}_{\rm sat}}{V_{\rm sat}^3}\nonumber\\
&&\times\left[{\rm erf}(X_v)-\frac{2X_v}{\sqrt{\pi}}\exp\left(-X_v^2\right)\right]\, ,
\end{eqnarray}
where $\mathbf{V}_{\rm sat}$ is the satellite velocity, $r_{\rm sat}$ is the satellite position within the host halo, $X_v=V_{\rm sat}/\sqrt{2}\sigma_v$, $\sigma_v$ is the velocity dispersion of the host halo, and $\ln\Lambda$ is the Coulomb logarithm.  For $\rho_{\rm host}$ we assume the density profile of \citet{1997ApJ...490..493N} (NFW).\footnote{\textsc{Galacticus} supports multiple different profiles, including NFW and Einasto, and our new code will work with any of these.}  We also set the Coulomb logarithm to $\ln\Lambda=2.0$, which is similar to the value used by \cite{2001ApJ...559..716T}.  Also, since we do not estimate the merging time beforehand, we set a condition that the satellite is considered merged with the parent halo when
\begin{eqnarray} \label{E:merge}
r_{\rm sat}<{\rm max}(r_{1/2,{\rm host}}+r_{1/2,{\rm sat}},0.01r_{\rm vir})\, ,
\end{eqnarray}
where $r_{1/2,{\rm host}}$ and $r_{1/2,{\rm sat}}$ are the half-mass radii of the host and satellite (subhalo) galaxies, respectively, and $r_{\rm vir}$ is the virial radius of the host halo.

\subsection{Tidal Stripping}
Satellite mass loss occurs due to gravitational tidal forces from the host halo.  To a first approximation, all the satellite mass outside the tidal radius $x_t$ is stripped over the course of an orbit \citep{2007MNRAS.374..775K,2010MNRAS.402.1899K}. The tidal radius $x_t$ is given by
\begin{eqnarray}\label{E:rtidal}
x_t=\left(\frac{GM_{\rm sat}(<x_t)}{\omega^2-d^2\Phi/dr^2}\right)^{1/3}\, ,
\end{eqnarray}
\citep{1962AJ.....67..471K} where $M_{\rm sat}(<x_t)$ is the enclosed mass within the tidal radius of the satellite halo, $\omega$ is the angular velocity of the satellite, and $\Phi$ is the gravitational potential from the host halo, such that
\begin{eqnarray}
\left.\frac{d^2\Phi}{dr^2}\right|_{r=r_{\rm sat}}=-\frac{2GM_{\rm host}(<r_{\rm sat})}{r_{\rm sat}^3}+4\pi G\rho_{\rm halo}(r_{\rm sat})\, ,
\end{eqnarray}
where $M_{\rm host}(<r)$ is the enclosed mass within radius $r$ of the host halo.

We use Eq.~8 of \citet{2005ApJ...624..505Z}, which converts the mass loss over a time interval to a mass loss rate, given by
\begin{eqnarray}
\frac{dM_{\rm sat}}{dt}=-\alpha\frac{M_{\rm sat}(>x_t)}{T_{\rm orb}}\, ,
\end{eqnarray}
where $T_{\rm orb}=2\pi/\omega$ is the orbital period of the satellite, $M_{\rm sat}(>x_t)$ is the satellite mass outside the tidal radius, and $\alpha$ is a parameter that absorbs the complicated details of tidal mass loss, including details dependent on the subhalo structure.  We set $\alpha=2.5$ by calibrating to N-body simulations (see Sec.~\ref{S:sim}).  We also set another disruption condition, in that the satellite is considered disrupted when it has lost 99\% of its initial mass.

\subsection{Tidal Heating}
Rapid gravitational encounters between the satellite and the parent halo can produce adiabatic tidal heating in the satellite.  This adiabatic heating expands the satellite, changing the density profile of the satellite and making it more susceptible to tidal stripping and merging.  In our analysis, we solve heating equations to determine the expansion of the half-mass radii, which increases the merger rate (see Eq.~\ref{E:merge}), and the \emph{contraction} of the tidal radius, which increases the mass loss.

We assume each subhalo starts with an initial thermal (or kinetic) energy given by the virial theorem, and we derive the energy rate deposited by a gravitational encounter based on the formalism of \citet{1999ApJ...514..109G} and \citet{2001ApJ...559..716T}.  Note that this energy rate is \emph{per unit mass}, which is important in our subsequent derivation of the expansion.  It is shown in these papers that the first-order change in the satellite energy $\Delta E_1$ is simply the work done by the tidal acceleration,
\begin{eqnarray}
\Delta E_{1,{\rm imp}}(t) = W_{\rm tid}(t) = \frac{1}{2}\Delta V_{\rm tid}^2=\frac{1}{2}\left[\int_0^tdt'\,\vecA_{\rm tid}(t')\right]^2\, .
\end{eqnarray}
Since we evolve the differential equations directly, we convert this to an energy rate, given by
\begin{eqnarray}
\frac{dE_{1,{\rm imp}}}{{\rm d}t}=\frac{d\Delta E_{1,{\rm imp}}}{dt} = \vecA_{\rm tid}(t)\int_0^t{\rm d}t'\,\vecA_{\rm tid}(t')\, .
\end{eqnarray}
The tidal acceleration is given by
\begin{eqnarray}
\vecA_{\rm tid}(t)=\vecx\cdot[\nabla\vecg]_{(\vecx=0)}=g_{a,b}(t)x_b\mathbf{e}_a\, ,
\end{eqnarray}
where $\vecx(t)$ is the position of a satellite mass element relative to its center, $\mathbf{e}_a$ is the unit vector in the $x_a$ direction, $g_{a,b}$ is the tidal tensor evaluated at $\vecx=0$, and the repeated indices $a$, $b$ are summed over the Cartesian coordinates.  The tidal tensor is given by
\begin{eqnarray}
g_{a,b} &=& \frac{GM_{\rm host}(<r)}{r^3}\left(\frac{3r_ar_b}{r^2}-\delta_{a,b}\right)\nonumber\\
&&-4\pi G\rho_{\rm host}\frac{r_ar_b}{r^2}\, ,
\end{eqnarray}
where $\mathbf{r}=\mathbf{r}_{\rm sat}$ in this equation and $\delta_{a,b}$ is the Kronecker delta.  We can now insert the tidal acceleration into the energy rate, giving us
\begin{eqnarray}
\frac{{\rm d}E_{1,{\rm imp}}}{{\rm d}t}&=&g_{a,b}(t)x_b(t)\mathbf{e}_a\cdot \int_0^tdt'\,g_{c,d}(t')x_d(t')\mathbf{e}_c\nonumber\\
&=&x_b(t)x_d(t)g_{a,b}(t)\int_0^tdt'\,g_{a,d}(t')\, .
\end{eqnarray}
Note that we use the impulse approximation where $\Delta x/x\ll1$, allowing us to move $x_d(t)$ out of the integral.  We then average the energy rate over a sphere, allowing us to use the fact that
\begin{eqnarray}
\VEV{x_ax_b}=\frac{1}{3}x^2\delta_{a,b}\, ,
\end{eqnarray}
where $x=|\vecx|$ to set
\begin{eqnarray}
\frac{{\rm d}E_{1,{\rm imp}}}{{\rm d}t}=\frac{1}{3}x^2g_{a,b}(t)G_{a,b}(t)\, ,
\end{eqnarray}
where
\begin{eqnarray}
G_{a,b}(t)=\int_0^t{\rm d}t'\,g_{a,b}(t')\, .
\end{eqnarray}

We implement corrections to this formula corresponding to those used in \citet{2001ApJ...559..716T}, namely accounting for the breakdown of the impulse approximation when the orbital and shock timescales are comparable and accounting for higher-order heating effects.  Implementing these corrections gives us an energy rate of the form
\begin{eqnarray}\label{E:dedt}
\frac{dE}{dt}=\frac{\epsilon_h}{3}\left[1+\left(\frac{T_{\rm shock}}{T_{\rm orb}}\right)^2\right]^{-\gamma}x^2g_{a,b}(t)G_{a,b}(t)\, .
\end{eqnarray}
The bracketed factor is the adiabatic correction discussed in \citet{1999ApJ...513..626G} with adiabatic index $\gamma=2.5$ and the shock timescale $T_{\rm shock}=r_{\rm sat}/v_{\rm sat}$.  The parameter $\epsilon_h$ is a heating coefficient that accounts for the higher-order heating effects.  We set $\epsilon_h=3$ by calibrating to N-body simulations (see Sec.~\ref{S:sim}).  Note that we do not evolve the energy directly. We use the impulse approximation, neglecting the time-variability of $x^2$ in Eq.~\ref{E:dedt}, and just track $Q=E/x^2$, which is independent of position within the satellite.

We now derive the radial expansion of a satellite mass element due to an energy transferred to the satellite through tidal heating.   We assume mass conservation and no shell crossings, allowing us to treat the satellite as a sum of spherical shells.  Assuming virial equilibrium, we can use the virial theorem to assert that the energy per unit mass of a spherical shell of radius $x$ is $\VEV{E(x)}=\frac{1}{2}\Phi(x)$, where
\begin{eqnarray}
\Phi(x)=-\frac{GM_{\rm sat}(<x)}{x}\, .
\end{eqnarray}
When energy $E(x)$ is injected into the shell, we use conservation of energy
\begin{eqnarray}
-\frac{GM'_{\rm sat}(<x_f)}{2x_f}=-\frac{GM_{\rm sat}(<x_i)}{2x_i}+E(x_i)\, ,
\end{eqnarray}
where $x_i$ and $x_f$ are the initial and final radii of the shell.  Mass conservation and no shell crossings allows us to set $M'_{\rm sat}(<x_f)=M_{\rm sat}(<x_i)$.  We use this and $E(x)=x^2Q$ to simplify our energy equation to the form
\begin{eqnarray}\label{E:expand}
\frac{1}{x_f}=\frac{1}{x_i}-\frac{2x_i^2Q}{GM_{\rm sat}(<x_i)}\, .
\end{eqnarray}

This equation allows us to track the expansion of a satellite mass element with heating.  For the merging condition, we expand the half-mass radius for the satellite based on Eq.~\ref{E:expand}.  For tidal stripping, we replace the final (post-expansion) tidal radius on the left-hand side of Eq.~\ref{E:rtidal} with its initial (pre-expansion) form using Eq.~\ref{E:expand} and solve for the initial tidal radius, which we then use to find the outer mass $M_{\rm sat}(>x_i)$.

We also use this formalism to determine the final internal density profile of subhalos.  In order to determine the density at radius $x_f$, we divide the subhalo into spherical shells and determine the density of each shell.  This gives each shell the density
\begin{eqnarray}
\rho'_{\rm sat}(x_f)=\frac{\Delta M_{\rm shell}}{4\pi x_f^2\Delta x_f}\, ,
\end{eqnarray}
where $\Delta x$ is the shell's width.  Using mass conservation, we write the shell's mass in terms of the initial halo parameters as $\Delta M_{\rm shell}=4\pi x_i^2\Delta x_i\rho(x_i)$, allowing us to write the final halo distribution in terms of the initial distribution as
\begin{eqnarray}
\rho'_{\rm sat}(x_f)=\left(\frac{x_i}{x_f}\right)^2\frac{{\rm d}x_i}{{\rm d}x_f}\rho_{\rm sat}(x_i)\, ,
\end{eqnarray}
where $x_i(x_f)$ is given by Eq.~\ref{E:expand}, and ${\rm d}x_i/{\rm d}x_f=({\rm d}x_f/{\rm d}x_i)^{-1}$ is found by differentiating both sides of Eq.~\ref{E:expand} and multiplying by $-x_f^2$, giving us
\begin{eqnarray}
\frac{{\rm d}x_f}{{\rm d}x_i}&=&\left(\frac{x_f}{x_i}\right)^2+\frac{4x_f^2x_i}{GM_{\rm sat}(<x_i)}\nonumber\\
&&-\frac{2x_f^2\Delta E(x_i)}{GM_{\rm sat}^2(<x_i)}\frac{{\rm d}M_{\rm sat}(<x_i)}{{\rm d}x_i}\, ,
\end{eqnarray}
which, using ${\rm d}M_{\rm sat}=4\pi x^2\rho_{\rm sat} {\rm d}x$, is
\begin{eqnarray}
\frac{{\rm d}x_f}{{\rm d}x_i}&=&\left[1-\frac{2x_i^3Q}{GM_{\rm sat}(<x_i)}\right]^{-2}\left[1+\frac{4x_i^3Q}{GM_{\rm sat}(<x_i)}\right.\nonumber\\
&&\left.-\frac{8\pi x_i^6Q}{GM_{\rm sat}^2(<x_i)}\rho_{\rm sat}(x_i)\right]\, .
\end{eqnarray}
This makes the final expression
\begin{eqnarray}
\rho'_{\rm sat}(x_f)&=&\left[1-\frac{2x_i^3Q}{GM_{\rm sat}(<x_i)}\right]^4\left[1+\frac{4x_i^3Q}{GM_{\rm sat}(<x_i)}\right.\nonumber\\
&&\left.-\frac{8\pi x_i^6Q}{GM_{\rm sat}^2(<x_i)}\rho(x_i)\right]^{-1}\rho_{\rm sat}(x_i)\, .
\end{eqnarray}


\subsection{Aquarius Calibration}\label{S:sim}
We calibrate the CDM subhalo abundance to the results from the Aquarius simulations \citep{2008MNRAS.391.1685S} over the mass range $10^8<M_{\rm sat}<10^{12}M_\odot$.  We set the mass resolution of our merger trees to $M_{\rm res}=5\times10^7M_\odot$.  We run \textsc{Galacticus} for Milky Way-sized parent halo masses $8\times10^{11}<M_{\rm host}<2\times10^{12}M_\odot$, the same range of parent masses used in the Aquarius simulations, with 800 trees/decade in mass.  We also match the cosmological parameters in our calculation with those in the simulations; specifically, we set $\Omega_m=0.25$, $\Omega_\Lambda=0.75$, $\sigma_8=0.9$, $n_s=1$, and $h=0.73$.  When calculating the abundance, we only include subhalos with a distance $r<r_{50}=433$ kpc, which is the value set for the Aquarius simulations.  When a halo first becomes a subhalo, we initialize the concentration of its NFW density profile according to the prescription of \citet{2008MNRAS.387..536G}, and we select orbital parameters for the subhalo using the distribution from \citet{2005MNRAS.358..551B}, assuming subhalo infall is isotropic.

\begin{figure}
\begin{center}
{\scalebox{0.45}{\includegraphics{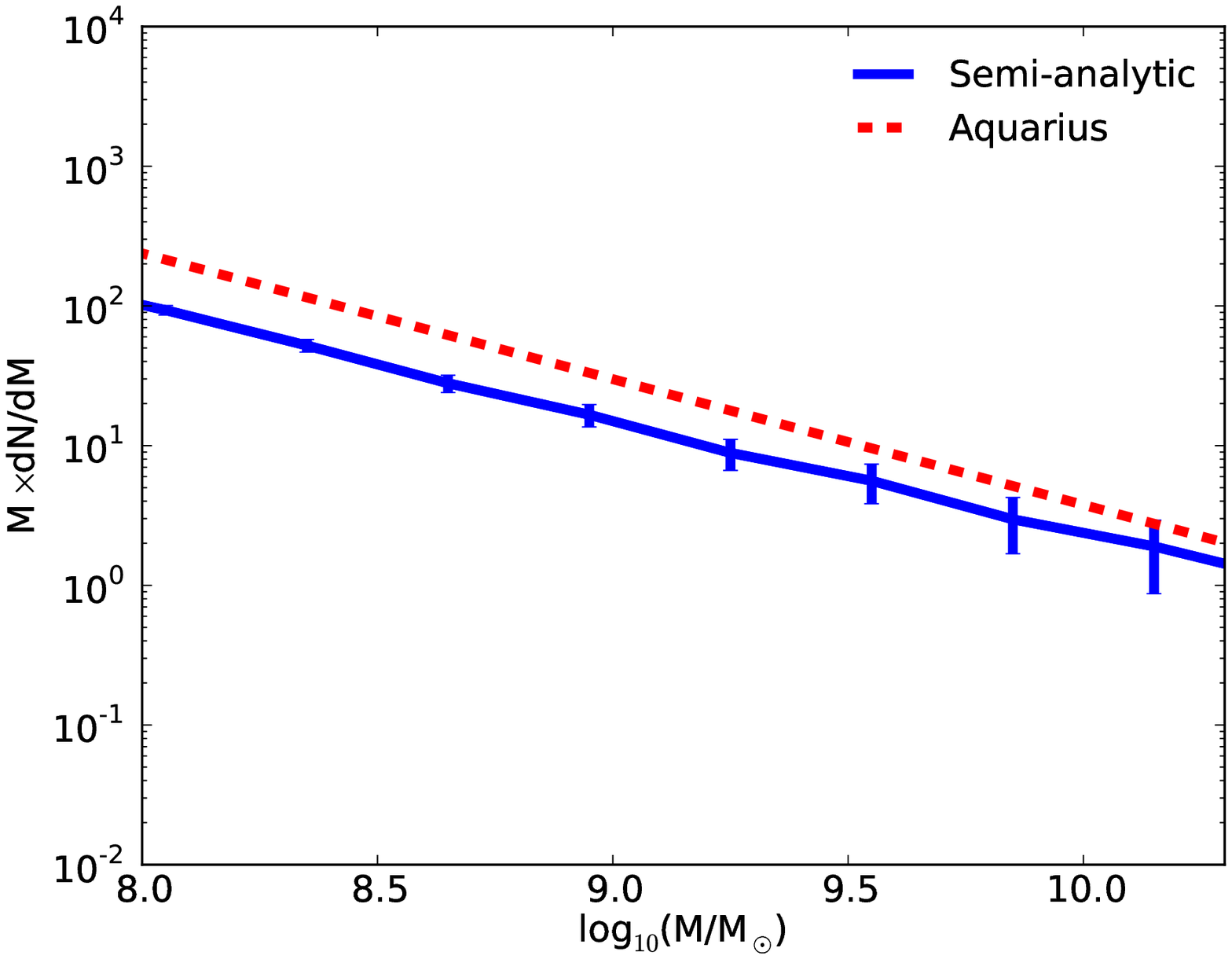}}}
{\scalebox{0.45}{\includegraphics{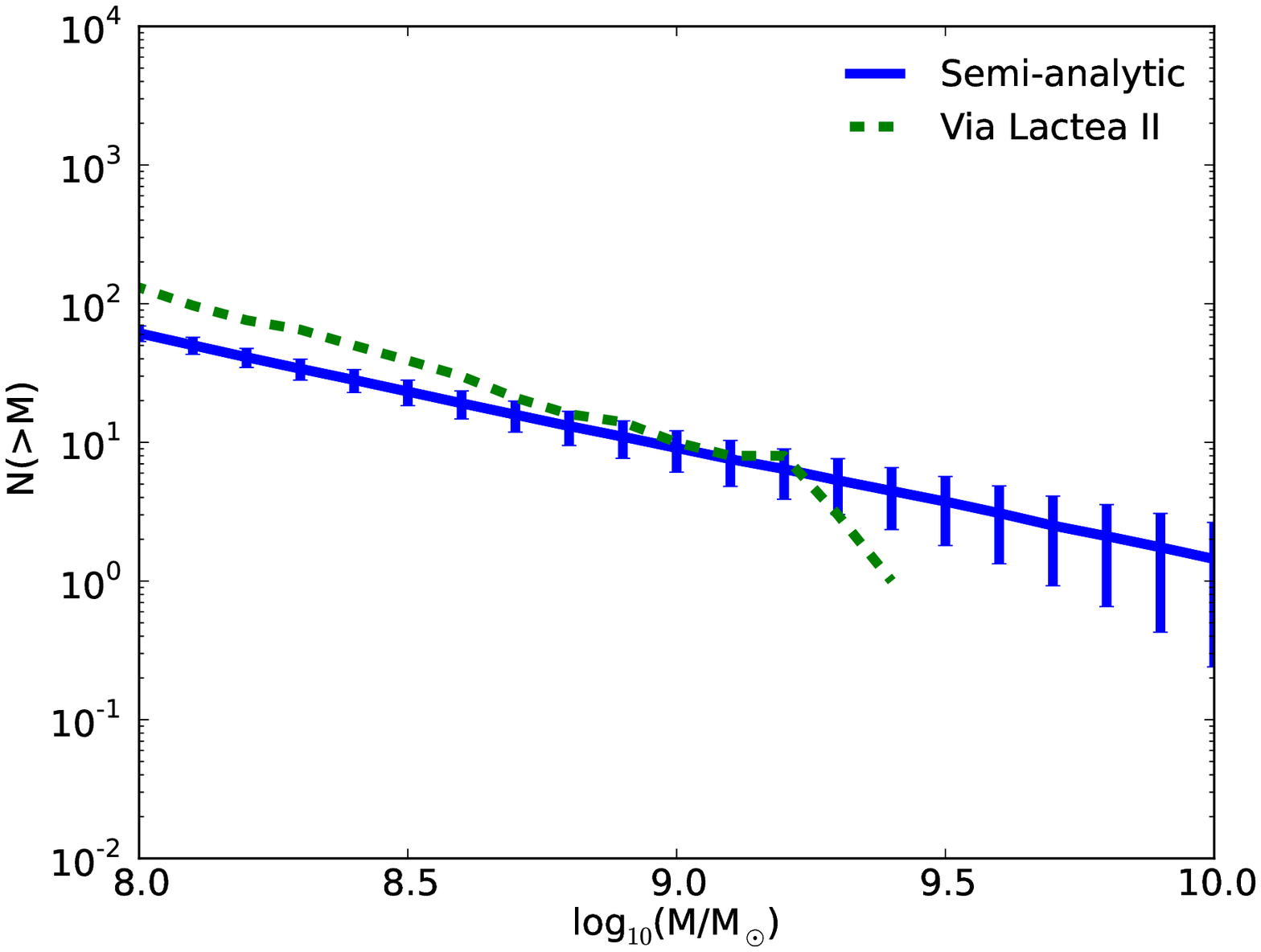}}}
\caption{\label{F:compsim} \emph{Upper panel}:The differential subhalo abundance from our CDM semi-analytic model (solid) with $\alpha=2$ and $\epsilon_h=3$ and the abundance of N-body simulation from the Aquarius simulations (dashed).  The error bars assume Poisson errors based on the total number of merger trees times $\sqrt{N_{\rm tree}/N_{\rm sim}}$, where $N_{\rm sim}=6$ is the number of Aquarius simulations.  \emph{Lower Panel}:The cumulative subhalo abundance from the same model and parameters as above (solid) and the abundance of N-body simulation from the Via Lactea II simulation (dashed).  In this case there was only one simulation, so the Poisson errors are multiplied by $\sqrt{N_{\rm tree}}$.}
\end{center}
\end{figure}

The results, with the mass loss parameter $\alpha=2$ and the heating coefficient $\epsilon_h=3$, are shown in Fig.~\ref{F:compsim}.  Our subhalo abundance is well-calibrated at high masses, yet it under-predicts the abundance at lower masses.  We increase the Poisson error bars, based on the number of merger trees, in this figure such that they are representative of the errors expected if just 6 merger trees were used to construct the mass function. This is to account for the fact that Aquarius includes only 6 simulations while we average over 318 merger trees in our calculation---the errors bars should therefore provide a reasonable indication of the expected difference between our model and Aquarius.

We also perform a comparison with the cumulative subhalo abundance from the Via Lactea II (VL-II) simulation \citep{2008Natur.454..735D}, shown in Fig.~\ref{F:compsim}.  In this case, we match the host mass in the simulation by running \textsc{Galacticus} for parent halo masses $10^{12}<M_{\rm host}<3\times10^{12}M_\odot$.  The subhalo abundance for this simulation included subhalos for $r<400$ kpc, and we impose the same condition for our tabulated abundance.  The cosmological parameters were also slightly different; we set $\Omega_m=0.238$, $\Omega_\Lambda=0.762$, $\sigma_8=0.74$, $n_s=951$, and $h=0.73$.  Similar to the previous simulation, we increase our Poisson errors such that they are equivalent to one merger tree, in order to correspond to there being only one simulation in VL-II.  In our comparison, we still under-predict the subhalo abundance at lower masses.  Note that the drop-off in the VL-II abundance for $\log_{10}(M/M_\odot)\gtrsim9.3$ is just an artifact due to small-number statistics.

For a first pass, we do see general agreement, and we acknowledge a few issues with our work.  First, we did not perform a full search in parameter space, \emph{i.e.} a \emph{Markov Chain Monte Carlo} (MCMC), for the nonlinear model parameters, in addition to not varying the Coulomb logarithm $\ln\Lambda$ for dynamical friction and the adiabatic index $\gamma$ for tidal heating.  We plan to perform a full MCMC for all the parameters in future work.  Second, our nonlinear model is imperfect, independently of the model parameters, in that various approximations are used.  Third, the difference in the results may be partially due to the difficulty of assigning halo masses in simulations, which has been explored by \citet{2014MNRAS.441.3488A}, as well as finding subhalos in simulations \citep{2014arXiv1403.6827J,2014arXiv1403.6835V}.  Fourth, N-body simulations can exhibit significant halo-to-halo variations in the subhalo abundance due to differences in the halo formation times, as shown in \citet{2009ApJ...696.2115I}.  These considerations need to be addressed in future work to forecast upcoming observations; however, our semi-analytic calculation is in general agreement with the Aquarius simulations, which allows us to be confident that we are able to produce a general picture of halo abundances for CDM halos to compare with WDM halos.


\subsection{Nonlinear Evolution Components}
\begin{figure}
\begin{center}
{\scalebox{0.45}{\includegraphics{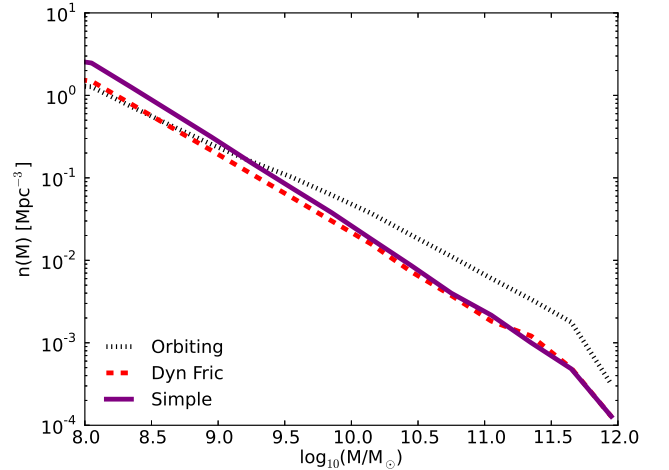}}}
\caption{\label{F:standard} The subhalo mass function from our CDM semi-analytic model including no nonlinear evolution (dotted) and only dynamical friction (solid) and the same from the simple scheme (dashed).}
\end{center}
\end{figure}

Next, we compare the ``simple'' satellite evolution scheme, where merging times are based on N-body simulations from \citet{2008ApJ...675.1095J} with no explicit tidal stripping modeling, with our ``orbiting'' scheme, where we follow the satellite orbits to determine the merging times dynamically.  We run merger trees for parent halos in the mass range $10^{12}<M_{\rm sat}<3\times10^{12}M_\odot$.  In Fig.~\ref{F:standard} we plot the subhalo mass functions from the simple scheme (solid curve), as well as the orbiting scheme with (dotted curve) and without (dashed curve) dynamical friction.  The subhalo mass function for the orbiting scheme is higher without dynamical friction because the satellites rarely merge into the host halo if there are no subhalo decelerations.  Dynamical friction increases with satellite mass, creating a larger effect at large masses.  The simple scheme and the orbiting scheme with dynamical friction, where satellites lose momentum causing them to be captured by the host halo, match well at large masses, suggesting the reliability of our orbiting scheme. However, these two schemes differ at low masses. This may be due to the fact that the merger timescales fitting formula from \citet{2008ApJ...675.1095J} used in the simple scheme was only calibrated to satellite masses $M_{\rm sat}> 10^{-2}M_{\rm host}$, where the two schemes do agree.  While our orbiting scheme is more complicated than the simple scheme of \citet{2008ApJ...675.1095J}, our scheme will provide more detailed predictions for subhalo populations.

\begin{figure}
\begin{center}
{\scalebox{0.45}{\includegraphics{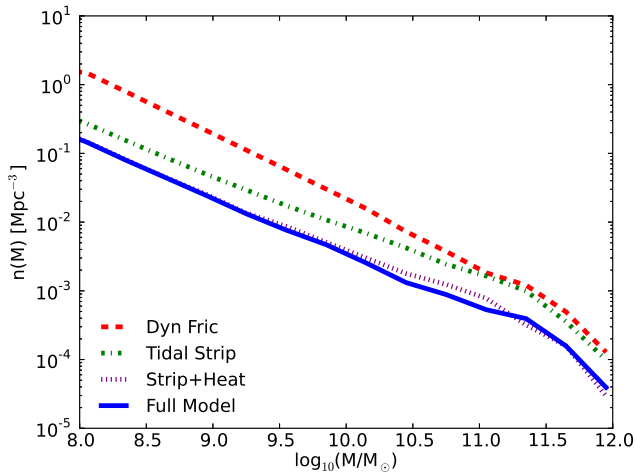}}}
\caption{\label{F:tidal} The subhalo mass function from our CDM semi-analytic model including only dynamical friction (dashed), only tidal stripping (dash-dotted), tidal stripping and tidal heating (dotted), and our full nonlinear model (solid).}
\end{center}
\end{figure}

Next, we include tidal effects. In Fig.~\ref{F:tidal} we add tidal stripping and tidal heating, as well as the complete subhalo mass function with all nonlinear evolution effects included.    Tidal stripping and tidal heating each have large effects on the subhalo mass function, particularly at low masses, confirming the importance of tidal effects to subhalo population evolution.  In particular, once tidal effects are included, adding dynamical friction does not affect the subhalo population much, presumably because the more massive halos that are susceptible to dynamical friction have already lost much mass due to tidal stripping.

\subsection{Warm Dark Matter}

\cite{2001ApJ...558..482B} consider the WDM linear theory power spectrum, $P(k)$. For $z=0$, they use the form suggested by \cite{2001ApJ...556...93B} which utilizes a fitting function for $T_{\rm WDM}(k)$. For $z=0$ this fitting function has parameters $\epsilon=0.361$; $\eta=5.00$; $\nu=1.2$; $R_{\rm c}= R_{\rm c}^0$. \cite{2001ApJ...558..482B} also derive their own form (using a Boltzmann code) for $T_{\rm WDM}(k)$ at $z=z_{\rm eq}$ (matter-radiation equality) and find $\epsilon=0.359$; $\eta=3.81$; $\nu=1.1$; $R_{\rm c}=0.932 R_{\rm c}^0$. The main difference in $T_{\rm WDM}(k)$ in $\eta$, which is reasonable as it controls the cut-off in the power spectrum which should be less sharp at $z=z_{\rm eq}$ than at $z=0$.

In \cite{2013MNRAS.428.1774B} we used the parameters from \cite{2001ApJ...556...93B} appropriate for $z=0$. In this work we instead use parameters appropriate for $z=z_{\rm eq}$ from \cite{2001ApJ...558..482B} as we explicitly take into account the effects of WDM at lower redshift by modifying the collapse barrier as described below.

To account for the evolution of perturbations into the non-linear regime, we follow the methodology of \cite{2001ApJ...558..482B}. The important consideration here is the velocity dispersion, $v_{\rm rms}$, of the WDM particles. This always scales as $v_{\rm rms}(z) = (1+z) v_{\rm rms}(z=0)$. \cite{2001ApJ...558..482B} define a ``Jeans mass'' for WDM corresponding to the linear regime at the initial time for their spherical collapse calculations. They then perform 1-D hydrodynamical simulations beginning from $z_{\rm eq}$ so that they can capture the effects of free-streaming which are mostly important at high-$z$. They report the resulting critical linear theory overdensity for collapse, $\delta_{\rm c}$, at $z=6$.

\cite{2001ApJ...558..482B} specifically discuss the appropriate way to include these velocities in a semi-analytic approach, noting that they obtain the closest match to numerical simulations if their calculation of the critical overdensity for collapse is begun at the redshift of equality. 
Therefore, to implement WDM semi-analytically, we use $P(k|z=z_{\rm eq}) = P_{\rm CDM}(k)  T_{\rm WDM}(k,z_{\rm eq})^2$, and $\delta_{\rm c}(z;M)$ taken from \cite{2001ApJ...558..482B}.

Two specific aspects of this approach warrant further clarification. First, it may seem that this method is effectively ``double counting'' the effects of WDM by including both a suppression in $P(k)$ and also a modified collapse barrier. In fact, this is not double counting. \cite{2001ApJ...558..482B} use linear theory to follow the WDM structure growth up to $z_{\rm eq}$---that is, they use $P_{\rm CDM}(k) T_{\rm WDM}(k)^2$ to get the power spectrum of linear perturbations at $z=z_{\rm eq}$. They then use hydrodynamical simulations with initial temperature matched to the WDM velocity dispersion at $z=z_{\rm eq}$ to do the later evolution. To match this in the semi-analytics, we therefore need to use $P_{\rm WDM}(k;z_{\rm eq})$ as our power spectrum, along with $\delta_{\rm c}(z;M)$ from \cite{2001ApJ...558..482B} to capture both the suppression of growth before and after $z_{\rm eq}$. To put this another way, $\delta_{\rm c}(z;M)$ tells us how big the perturbation, linearly extrapolated from $z_{\rm eq}$ to $z$, has to be in order to collapse. But, $P(k)$ tells us the distribution of perturbation amplitudes at $z_{\rm eq}$ extrapolated to $z$.

A second consideration relates to our use of the \cite{2001ApJ...558..482B} barrier at all redshifts, while \cite{2001ApJ...558..482B} derived it specifically for $z=6$. It is important to note that our barrier takes the form of a redshift-independent multiplicative factor which scales $\delta_{\rm c,CDM}(z)$, which we fit to the $z=6$ results of \cite{2001ApJ...558..482B}. That this approach is reasonable can be understood by considering how modes of given wavenumber grow in a WDM universe:
\begin{itemize}
 \item The WDM free-streaming scale (comoving wavenumber), $k_{\rm fs}$, is constant after the WDM becomes non-relativistic, until the epoch of matter-radiation equality: $k_{\rm fs} = k_{\rm c,eq}$;

 \item After matter-radiation equality the free-streaming scale changes as the WDM velocities are redshifted away such that $k_{\rm fs} = k_{\rm c,eq} (a/a_{\rm eq})^{1/2}$.
\end{itemize}
 Consider then the growth of a mode of wavenumber $k$:
\begin{itemize}
  \item If $k < k_{\rm c,eq}$ then the mode is never affected by free streaming, so its linear growth function is just that of CDM, i.e. $D(a) = D_{\rm CDM}(a)$;

  \item If $k > k_{\rm c,eq}$ then there is an epoch, $a_c = a_{\rm eq} (k/k_{\rm c},eq)^2$, at which $k = k_{\rm fs}$:
\begin{itemize}
    \item For $a < a_c$ the mode has never been able to grow so $D(a) = D_{\rm CDM}(a_{\rm eq})$;

    \item For $a > a_c$ the mode has been able to grow since $a_c$, so $D(a) = D_{\rm CDM}(a)/D_{\rm CDM}(a_c)$.
\end{itemize}
\end{itemize}
Obviously there will be a smooth transition between these regimes---that will be accounted for by fitting to the detailed shape of the barrier determined by \cite{2001ApJ...558..482B}.  Here, all we want to do is establish the expected scalings. The critical overdensity for collapse should scale inversely with the growth function (the more the mode has been able to grow, the lower its initial overdensity could have been and it still undergoes collapse):
\begin{itemize}
 \item If $k < k_{\rm c,eq}$ then $\delta_{\rm c}(a) = \delta_{\rm c,CDM}(a)$  which is independent of $k$;

 \item If $k < k_{\rm c,eq}$ then:
 \begin{itemize}
    \item For $a < a_{\rm c}$ then $\delta_{\rm c}(a) = \delta_{\rm c,CDM}(a) D(a)/D(a_{\rm eq})$;

    \item For $a > a_{\rm c}$ then $\delta_{\rm c}(a) = \delta_{\rm c,CDM}(a) D(a_{\rm c})/D(a_{\rm eq}) \propto (k/k_{\rm c,eq})^2$ [assuming $D \propto a$]\footnote{This scaling is at least roughly in agreement with that found by \protect\cite{2001ApJ...558..482B}.}
\end{itemize}
\end{itemize}
 The case of modes that are still below the free-streaming scale is not of interest. Given the scaling $M_{\rm fs} \propto k_{\rm fs}^3 \propto a_{\rm c}^{3/2}$ then the mass scale for free-streaming at $z=0$ is roughly $4.6\times10^{-6}$ of that at equality, and the mass scale for free-streaming at $z=30$ is $7.6\times10^{-4}$ times that at equality. So, for any redshifts of interest these mass scales are so far below the WDM cut off introduced by free streaming that we can ignore this.

The above shows us that we expect $\delta_{\rm c}(a)$ to have a functional form that depends only on $k_{\rm c,eq}$ and which depends on expansion factor only through the usual CDM growth factor. This is exactly what we do with the \cite{2001ApJ...558..482B} determination of $\delta_{\rm c}$.

We now choose a test case for WDM, namely a particle mass $m_X=1.5$ keV and an effective number of degrees of freedom $g_X=1.5$.  As in Paper I, the transfer function $T(k)$ is modified for the WDM case to impose a cutoff below a free-streaming length scale $\lambda_s$ \citep{2001ApJ...556...93B,2001ApJ...558..482B}:
\begin{eqnarray}
T(k)\rightarrow T(k)\left[1+(\epsilon k\lambda_s)^{2\nu}\right]^{-\eta/\nu}\, ,
\end{eqnarray}
where $\epsilon=0.359$, $\eta=3.81$, and $\nu=1.1$.  For our test case, the cutoff scale is $\lambda_s=0.1125$ Mpc and the corresponding mass scale is $M_s=2.22\times10^8M_\odot$.   We also use the mass-dependent collapse threshold for WDM $\delta_{c,{\rm WDM}}$ given in Eq.~7 of Paper I, which affects the excursion set barrier.  We set the mass-concentration relation for WDM, by modifying it according to the prescription from \citet{2012MNRAS.424..684S}
\begin{eqnarray}
\frac{c_{\rm WDM}(M)}{c_{\rm CDM}(M)}=\left(1+\gamma_1\frac{M_{\rm hm}}{M}\right)^{-\gamma_2}\, ,
\end{eqnarray}
where $\gamma_1=15$ and $\gamma_2=0.3$ and $M_{\rm hm}=2.5\times10^9M_\odot$ is the half-mode mass scale, which corresponds to the length scale where the WDM transfer function is reduced by 1/2.

\begin{figure}
\begin{center}
{\scalebox{0.45}{\includegraphics{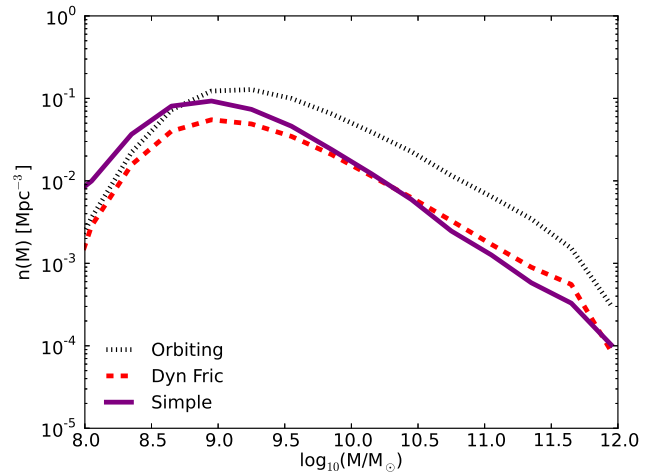}}}
\caption{\label{F:wdmstandard} The subhalo mass function from our WDM semi-analytic model including no nonlinear evolution (dotted) and only dynamical friction (solid) and the same from the simple scheme (dashed).}
\end{center}
\end{figure}

\begin{figure}
\begin{center}
{\scalebox{0.45}{\includegraphics{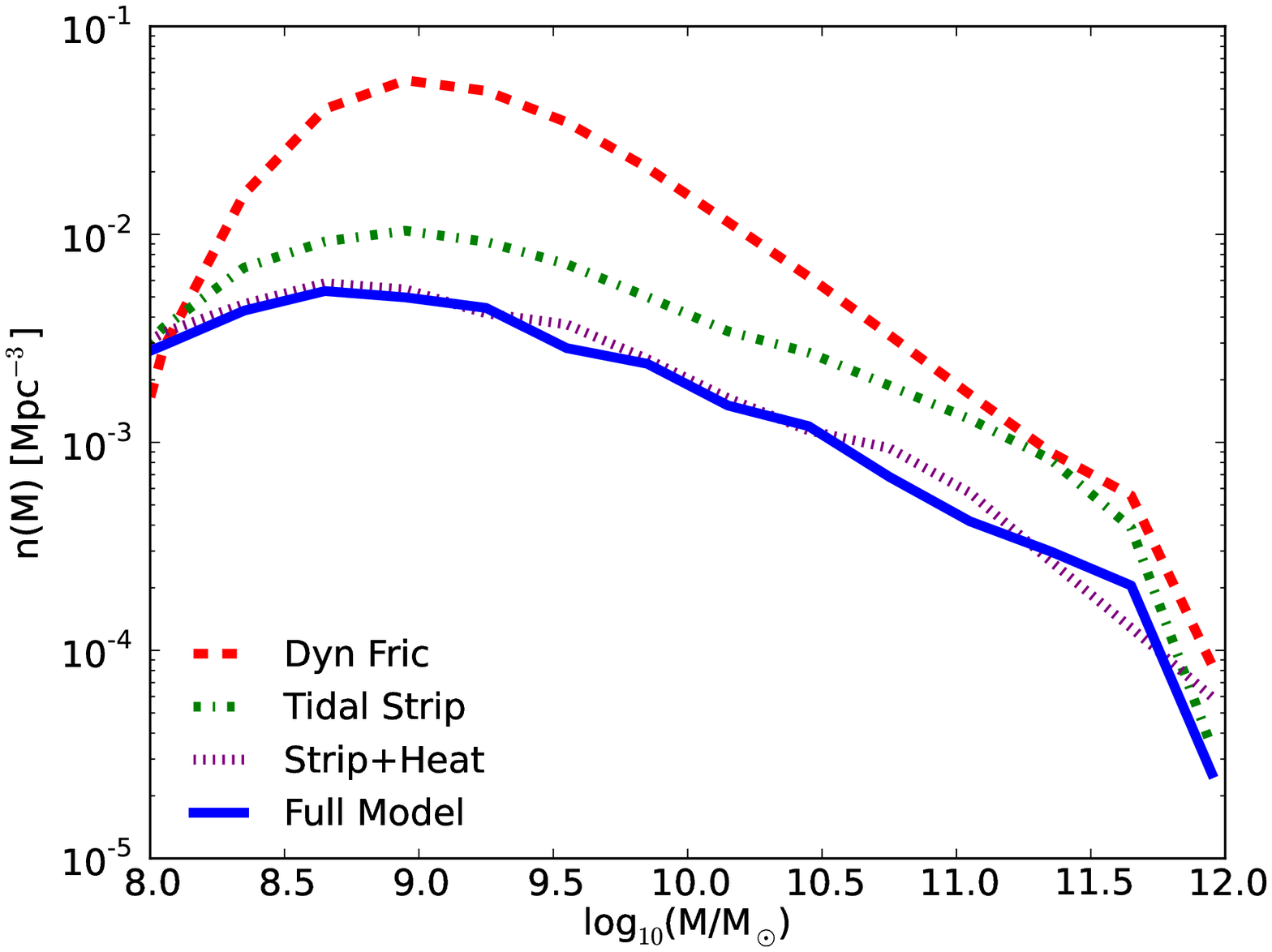}}}
\caption{\label{F:wdmtidal} The subhalo mass function from our WDM semi-analytic model including only dynamical friction (dashed), only tidal stripping (dash-dotted), tidal stripping and tidal stripping (dotted), and our full nonlinear model (solid).}
\end{center}
\end{figure}

\begin{figure}
\begin{center}
{\scalebox{0.45}{\includegraphics{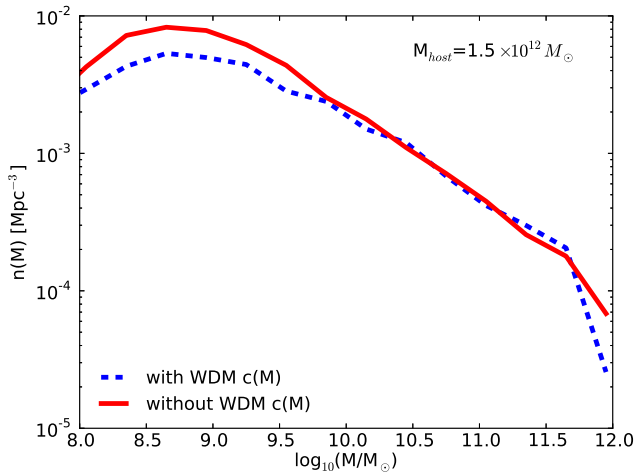}}}
\caption{\label{F:wdmconc} The subhalo mass function from our WDM semi-analytic model with(dashed) and without (solid) the CDM to WDM transformation to the mass-concentration relation from \citet{2012MNRAS.424..684S}. We see that the lower concentration in WDM halos causes them to be more susceptible to tidal effects.}
\end{center}
\end{figure}

Using the same semi-analytic nonlinear evolution prescription as before with the above prescriptions for WDM, we calculate the subhalo mass function for the WDM model.  Note that we use the same nonlinear evolution parameters as for CDM ($\alpha=2.5$, $\epsilon_h=3$).\footnote{The mass loss parameter $\alpha$ may change for other DM models, we expect it to be a small effect.  These parameters will be calibrated to WDM simulations in future work.} Similarly to the CDM case, in Fig.~\ref{F:wdmstandard} we plot mass functions using the simple scheme, as well as the orbiting scheme with and without dynamical friction. In Fig.~\ref{F:wdmtidal}, we add in tidal effects and show the full model.  The variations in the results for different nonlinear effects behave similarly to the CDM case, in that tidal effects dominate the subhalo mass function behavior.  In order to assess the effect of the WDM mass concentration prescription, we plot in Fig.~\ref{F:wdmconc}  the subhalo mass function with a CDM $c(M)$ (solid) and a WDM $c(M)$ (dashed).  We see that the lower concentration makes the halos significantly more susceptible to tidal stripping and heating---ignoring this change in concentrations would lead to the mass function being overestimated by around 50\% at its peak.

\section{CDM vs.~WDM Discussion} \label{S:cdmwdm}

This section compares the subhalo mass functions, radial distributions, and density profiles for CDM and WDM models evolved to the present day for Milky-Way-sized parent halos using our semi-analytic calculations.  We confirm the suppression of the WDM subhalo mass function at small masses.  We also show that our semi-analytically computed radial distributions are consistent with previous work.  Finally, we display radial distributions of subhalos and the resulting density profiles of the subhalos for both CDM and WDM models.

\begin{figure}
\begin{center}
{\scalebox{0.45}{\includegraphics{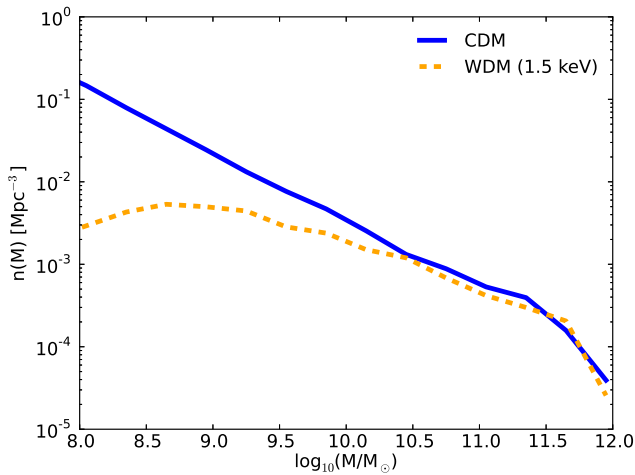}}}
\caption{\label{F:wdm} The subhalo mass function for CDM (solid) and WDM (dashed) from our semi-analytic model.}
\end{center}
\end{figure}

We plot the CDM and WDM subhalo mass functions for our semi-analytic model from the previous section in Fig.~\ref{F:wdm}.  The subhalo mass functions for the CDM and WDM models are coincident at large masses, but at smaller masses the WDM mass function is significantly suppressed.  This difference may be detectable in probes of sub-structure within lensing galaxies.

\begin{figure}
\begin{center}
{\scalebox{0.45}{\includegraphics{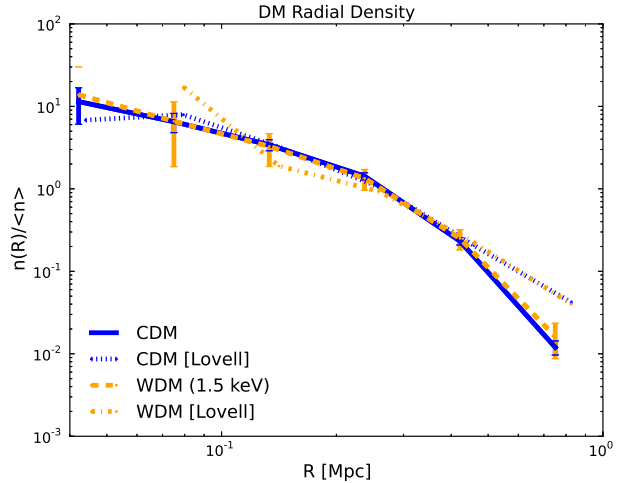}}}
\caption{\label{F:radlovell}Radial number densities of subhalos for semi-analytic CDM (solid) and WDM (dashed) models for subhalos with $M>10^8M_\odot$, normalized by the average number density over $r_{200b}=432.1$ (429.0) kpc for CDM (WDM), in rough agreement with corresponding CDM (dotted) and WDM (dash-dotted) results from \citet{2014MNRAS.439..300L}.  The error bars assume Poisson errors based on the total number of merger trees times $\sqrt{N_{\rm tree}/N_{\rm sim}}$, where $N_{\rm sim = 4}$ is the number of simulations in \citet{2014MNRAS.439..300L}}
\end{center}
\end{figure}

Before computing the radial number densities of subhaloes for different masses, we first compare the radial distribution based on our semi-analytic calculation with high-mass-resolution simulations from \citet{2014MNRAS.439..300L}.  In order to compare results, we determine the radial number density for all subhalos with masses $M>10^8M_\odot$, and then normalize these values to the average number density over their values of $r_{200b}$, the radius that encloses the region whereby the mean density is 200 times the background density.  The values for CDM (WDM) were $r_{200b}=432.1$ (429.0) kpc.  We plot our results for both DM models in Fig.~\ref{F:radlovell}, and find that our result is broadly consistent with Fig.~12 of \citet{2014MNRAS.439..300L}.  We do not detect a significant difference between the number densities for our CDM and WDM models, and the differences in the estimates for the CDM and WDM models in \citet{2014MNRAS.439..300L} are comparable to the error bars.  The differences between our models and \citet{2014MNRAS.439..300L} at small radii are marginally significant and will warrant attention when we calibrate our models in future work.

\begin{figure}
\begin{center}
{\scalebox{0.45}{\includegraphics{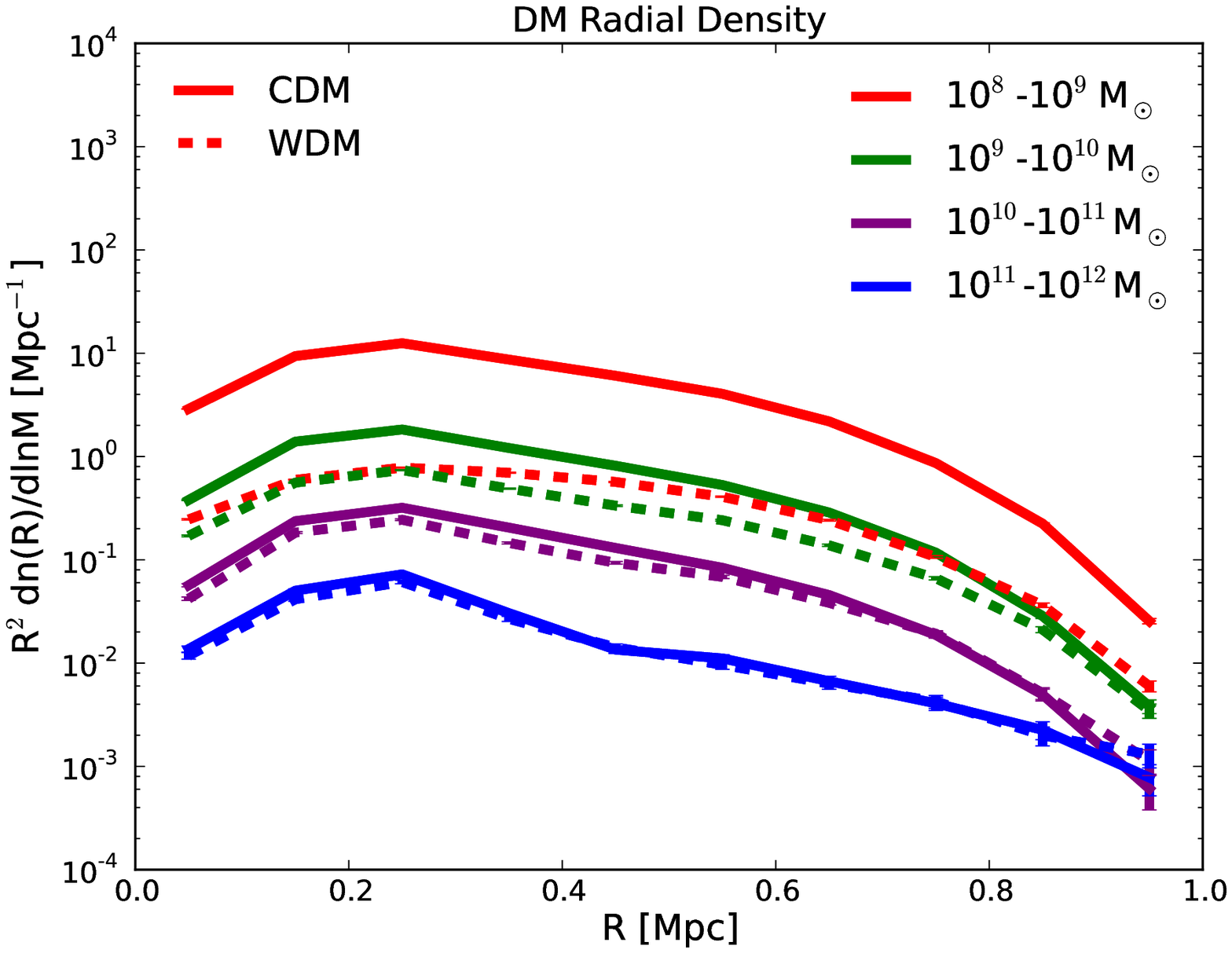}}}
{\scalebox{0.45}{\includegraphics{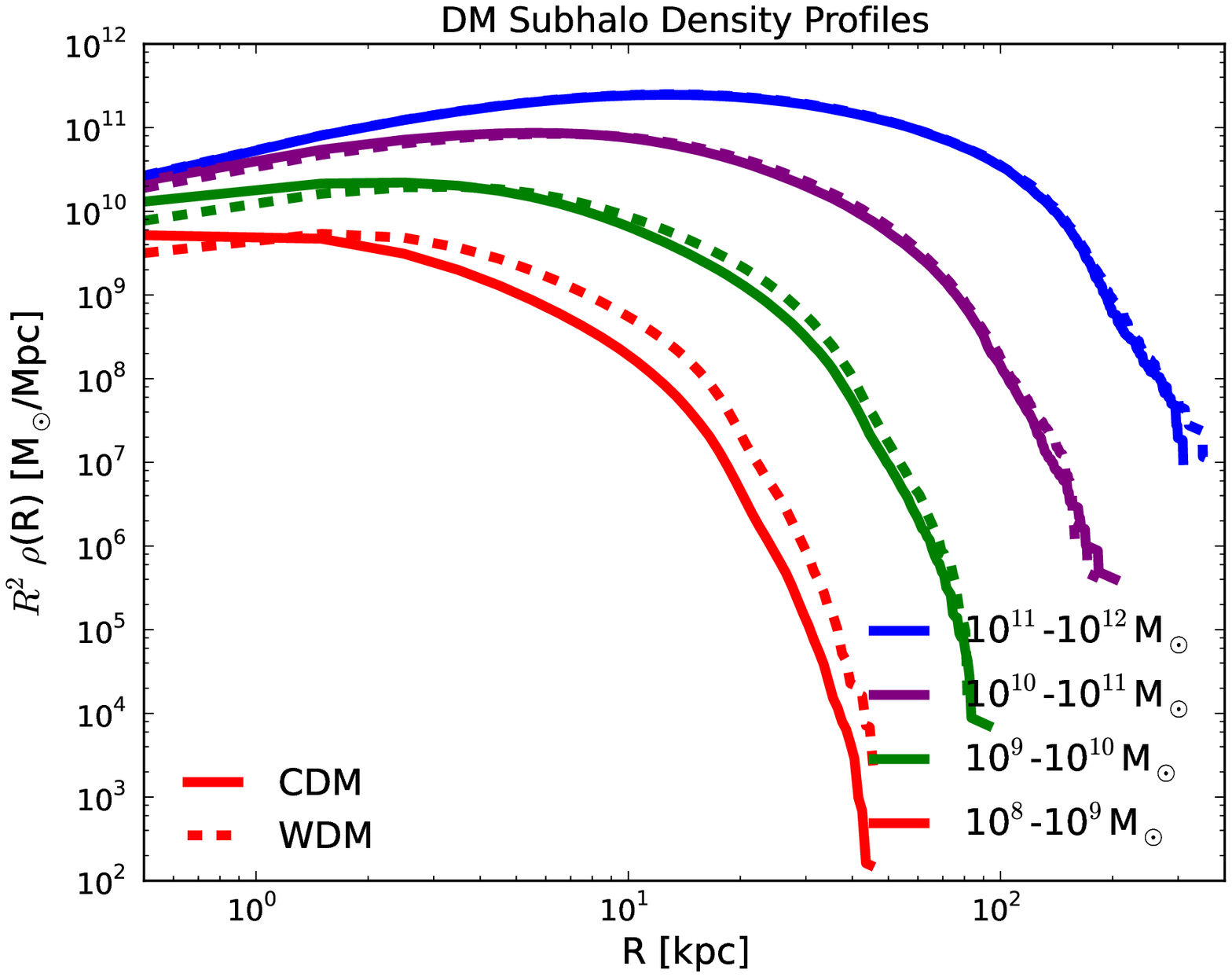}}}
\caption{\label{F:raddens} \emph{Upper panel}: Radial distributions of subhalos for CDM (solid) and WDM (dashed) models as a function of subhalo mass. The error bars denote Poisson errors based on the number of merger trees. \emph{Lower panel}: Density profiles of subhalos for CDM (solid) and WDM (dashed) models as a function of subhalo mass.}
\end{center}
\end{figure}

We then proceed to determine radial number densities for CDM and WDM models within specific mass ranges.  Fig.~\ref{F:raddens} shows the radial number density of subhalos relative to the host halo for the CDM and WDM models.  For these calculations, we increased the sampling to 19200 trees/decade in mass.  These results show that the overall radial number density tracks the subhalo mass function, in that for CDM the number density increases at low masses, while for WDM it is suppressed.  However, the number density profiles for all the mass ranges have similar variations with radius.  We find that CDM and WDM profiles have their greatest difference at small radii.

We also plot in Fig.~\ref{F:raddens} the mean final density profile of the subhalos for different mass ranges, assuming an NFW initial profile and our tidal heating formalism.  We see a small difference in the low-mass range between the CDM and WDM models, both at small and large radii within the subhalos.  The difference, however, is very small, requiring highly accurate predictions to be useful.  The CDM/WDM difference in the subhalo mass function is very large.  In future work, we will explore the relative traction in each of these probes for tracing the characteristics of dark matter.

Future constraints will rely on accurate predictions.  While our semi-analytic framework serves as a crucial step, a more accurate calibration to N-body simulations is necessary for robust and useful predictions that will enable discrimination between dark matter scenarios.


\section{Conclusions} \label{S:conc}
We implement nonlinear effects into semi-analytical modeling of subhalo orbits in order to more accurately predict the evolution of subhalo populations.  Specifically, the effects we considered include dynamical friction, tidal stripping, and tidal heating.  We calibrate our results for cold dark matter to N-body simulations, and we extend the calculations  to a test case warm dark matter model.  We also find that tidal effects dominate the nonlinear evolution, while dynamical friction only has a small effect. While the WDM subhalo mass function shows a significant suppression relative to CDM at lower masses, there are also small deviations between the models for radial subhalo distributions and subhalo density profiles.  We also show that the low concentration of WDM halos causes the halos to be slightly more susceptible to tidal stripping and heating.  We display differences in subhalo mass functions, radial distributions, and subhalo density profiles between CDM and WDM models that may be detectable in future lensing surveys.  We plan to apply this methodology to other models of dark matter, {\it i.e.} self-interacting dark matter, and to incorporate the effects of baryons (i.e. the presence of galaxies in both subhalos and host halos) on subhalo orbital evolution. With more accurate calibrations to N-body simulations, predictions of this kind could be used to help discriminate DM phenomenology in future lensing surveys.  This analysis can be repeated for various WDM parameter values and for other alternative dark matter models, \emph{i.e.}~self-interacting dark matter (SIDM), allowing a more robust test of dark matter phenomenology.

\begin{acknowledgments}

We thank T.~Abel, F.~Cyr-Racine, O.~Dor\'{e}, J.~Merten, and A.~Peter for helpful comments and useful discussions.  We also thank M.~Lovell and P.~Madau for providing various N-body simulation results.  Part of the research described in this paper was carried out at the Jet Propulsion Laboratory, California Institute of Technology, under a contract with the National Aeronautics and Space Administration. AP was supported by an appointment to the NASA Postdoctoral Program at the Jet Propulsion Laboratory, California Institute of Technology, administered by Oak Ridge Associated Universities through a contract with NASA.
\end{acknowledgments}

\end{document}